\documentclass[a4paper,10pt]{article}
\usepackage[utf8]{inputenc}

\usepackage[left=15mm,right=15mm,top=20mm,bottom=25mm]{geometry}
\usepackage{float}
\usepackage{wrapfig}
\usepackage[small,bf]{caption}
\usepackage{footnote}

\usepackage{graphicx}
\usepackage{amsmath}
\usepackage{amssymb}
\usepackage{bm}

\usepackage{tikz,xcolor,hyperref}

\definecolor{lime}{HTML}{A6CE39}
\DeclareRobustCommand{\orcidicon}{%
	\begin{tikzpicture}
	\draw[lime, fill=lime] (0,0) 
	circle [radius=0.16] 
	node[white] {{\fontfamily{qag}\selectfont \tiny ID}};
	\draw[white, fill=white] (-0.0625,0.095) 
	circle [radius=0.007];
	\end{tikzpicture}
	\hspace{-2mm}
}

\newcommand{\orcidSS}{\href{https://orcid.org/0000-0003-1677-8004}{\orcidicon}}
\newcommand{\orcidVP}{\href{https://orcid.org/0000-0002-3031-062X}{\orcidicon}}

\usepackage{fancyhdr}
\fancypagestyle{first}{%
  \fancyhf{}

  \fancyfoot[L]{{\footnotesize Preprint version created by the Authors in \LaTeX.}}
  \fancyfoot[R]{\thepage}
}

\fancypagestyle{header}{%
  \fancyhf{}

  \fancyhead[L]{Shore \& Pavl\'ik}
  \fancyhead[R]{\bf How a~fake Kepler portrait became iconic}
  \fancyfoot[L]{{\footnotesize Preprint version created by the Authors in \LaTeX.}}
  \fancyfoot[R]{\thepage}
}

\usepackage{multicol}
\title{How a~fake Kepler portrait became iconic}
\author{Steven N. Shore, Vaclav Pavlik}
\date{May 31, 2021}

\begin{document}
\pagestyle{header}
\thispagestyle{first}

% Figures

\begin{center}
	\section*{\Huge\textbf{How a~fake Kepler portrait became iconic}}

% Authors
	\textbf{Steven N.~Shore}\orcidSS $^1$ \&
	\textbf{V\'aclav Pavl\'ik}\orcidVP $^2$
	
	\smallskip
	
	\textit{%
		$^1$ Dipartimento di Fisica, Universit\`a di Pisa, Pisa, Italy\\
		$^2$ Astronomy Department, Indiana University Bloomington, Indiana, USA}
		
	\smallskip
	
	To appear in an edited form in Readers' Forum, Physics Today (September 2021)
\end{center}

\vspace{18pt}

\begin{multicols}{2}

% Text

\noindent
Textbooks and popular writings introduce portraits of historical personalities to illustrate the human side of science. Usually, they get it right. Einstein did stick out his tongue to reporters, Marie Curie really did dress in black and Oppenheimer wore a~fedora. But for the last few decades, one of the founders of modern physics and astronomy has been portrayed with a~fake. Since this year marks the 450$^{\rm th}$ anniversary of Johannes Kepler's birth, it is timely and necessary to point out an egregious example of unwittingly propagated misinformation.

\begin{figure*}[ht]
	\centering
	\textbf{a)} \includegraphics[height=6.8cm]{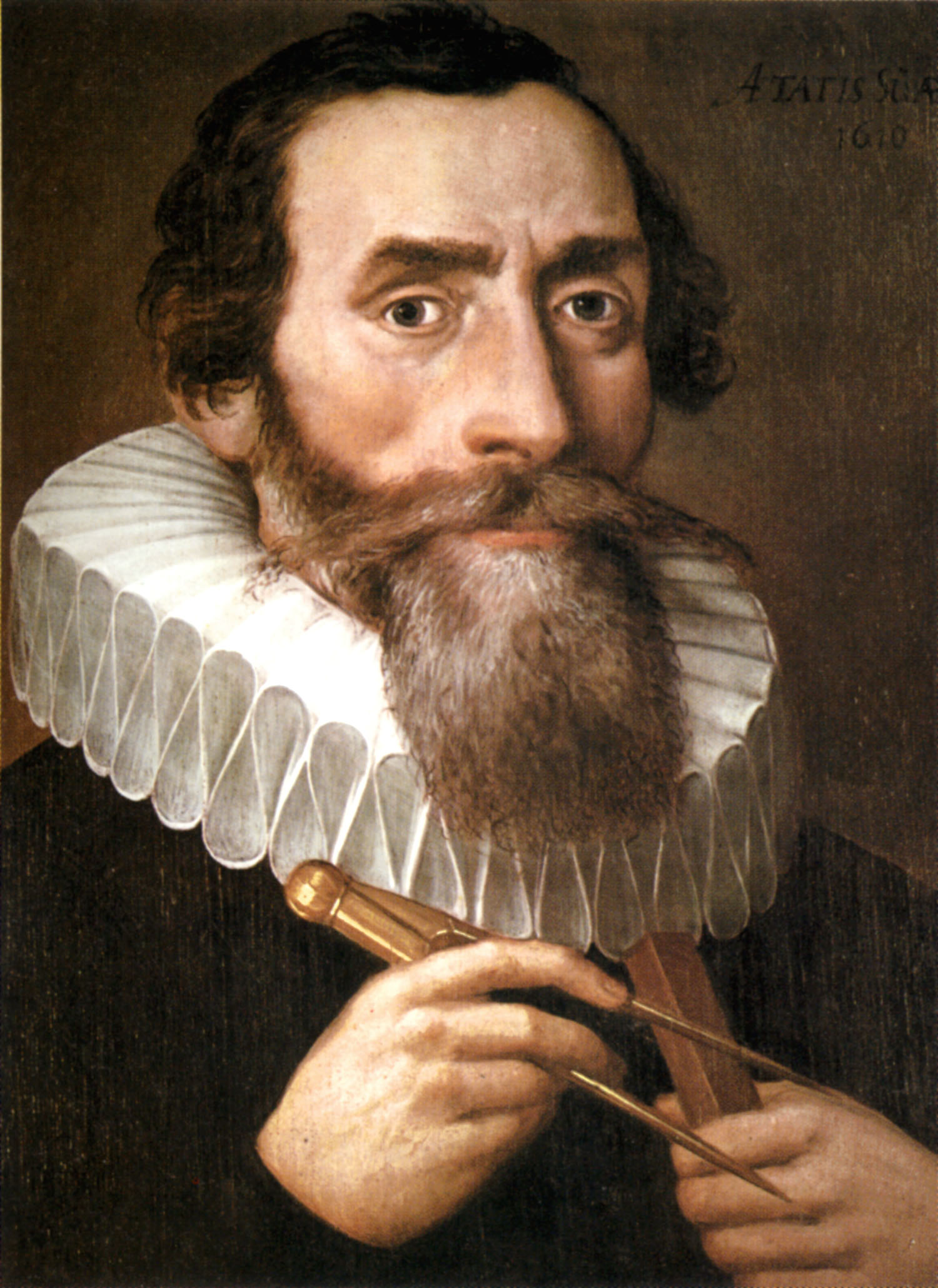}\hspace{2pt}
	\textbf{b)} \includegraphics[height=6.8cm]{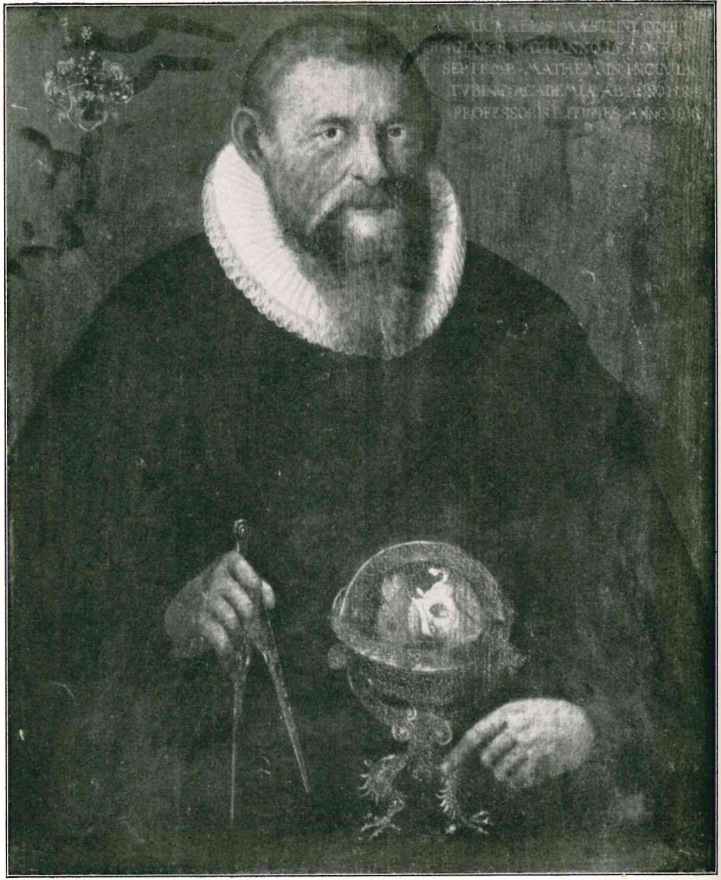}\\
	\vspace{10pt}
	\textbf{c)} \includegraphics[height=6.5cm]{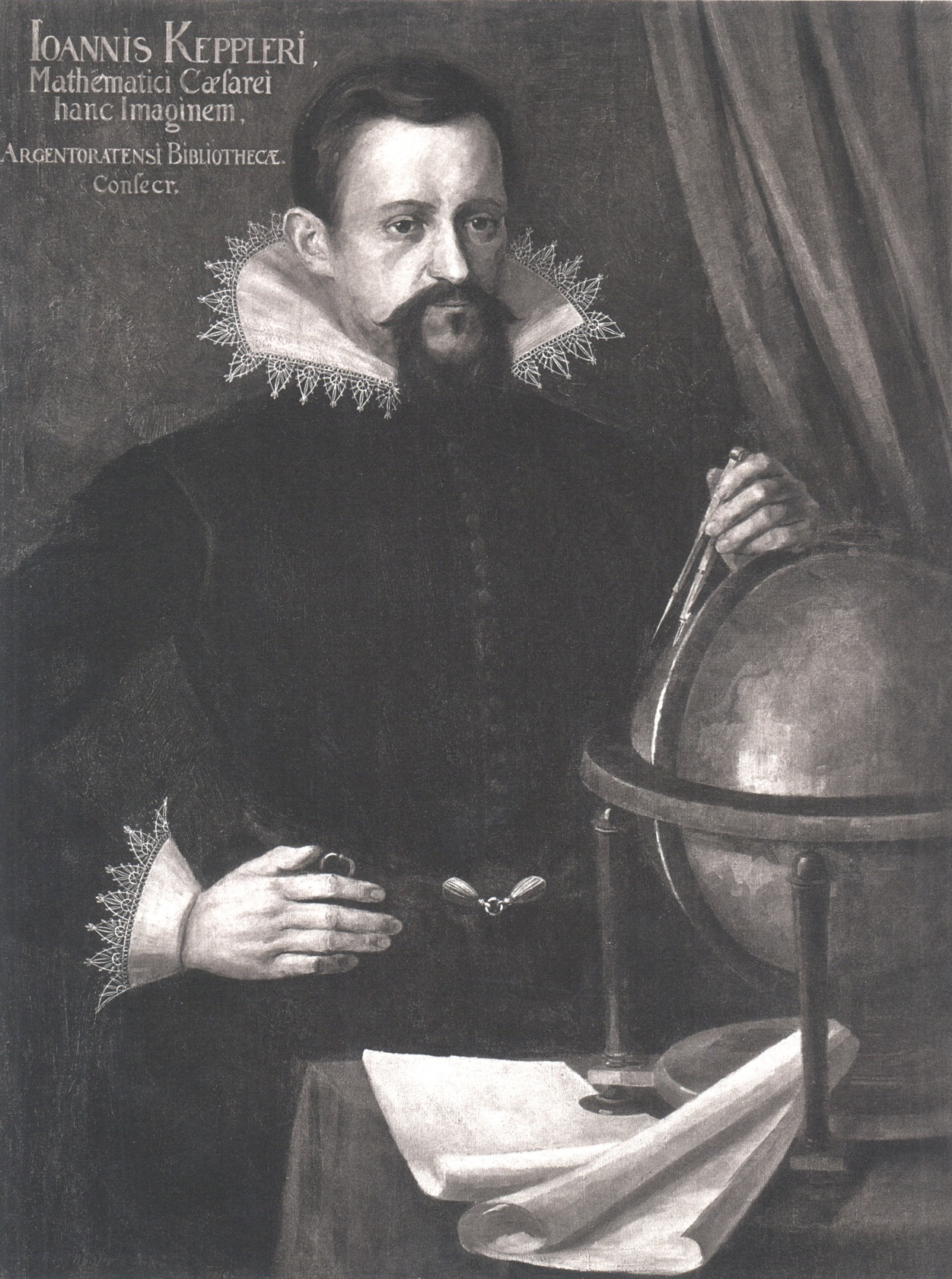}\hspace{2pt}
	\textbf{d)} \includegraphics[height=6.5cm]{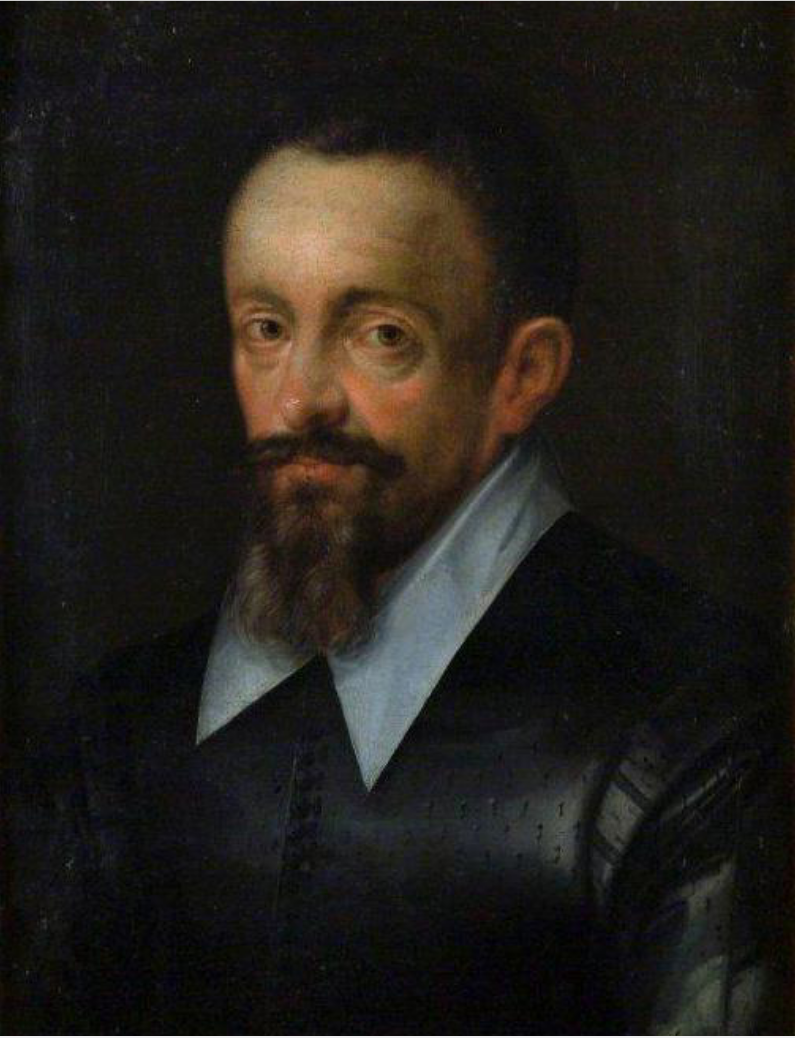} \hspace{2pt}
	\textbf{e)} \includegraphics[height=6.5cm]{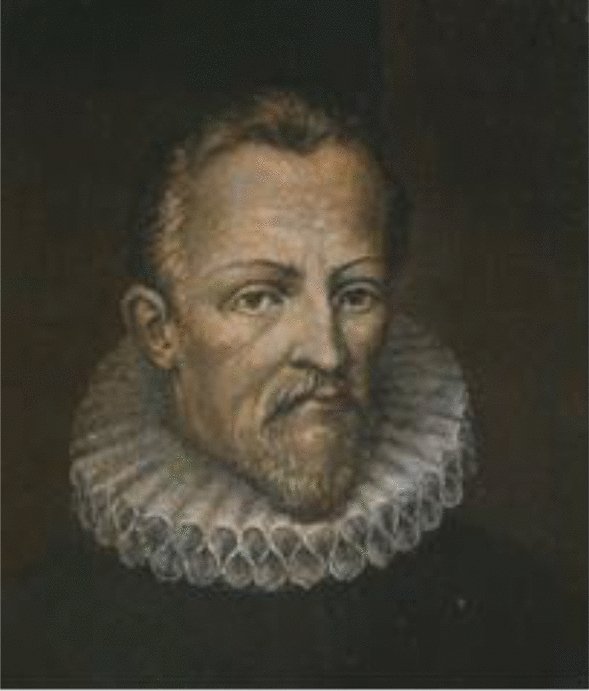} 
	
	\caption{\textbf{a) Fake Kepler portrait.} A~19$^{\rm th}$ century painting by an unknown artist, presumed to be a~copy of an unknown original, allegedly from 1610. \emph{If} it is based on anything, it more likely derived from a~portrait of Michael Mästlin.
	\textbf{b) Michael Mästlin portrait.} A~black and white photograph of an original from 1619 by an unknown artist, University of Tübingen, also flagged by Zinner \cite{zinner} as a~possible source for the fake.
	\textbf{c) Johannes Kepler portrait.} An engraving based on the 1620 Kepler portrait that was given to the Strasbourg library in 1627 (Courtesy of the Smithsonian Libraries and Archives, Image ID: SIL-SIL14-k001-08, \url{https://library.si.edu/image-gallery/72833}).
	\textbf{d) Presumed Kepler portrait.} Attributed to Hans von Aachen \cite{kaufmann}. It is assigned around the same year as the fake portrait, likely 1612.
	\textbf{e) The Linz miniature.} Presumed Kepler portrait from 1620, artist unknown \cite{ulm} (Courtesy of the Oberösterreichisches Landesmuseum, Linz).}
	\label{fig:kepler}
\end{figure*}

The portrait, Fig.~\ref{fig:kepler}a, that is now the first entry on Google search for ``Kepler portrait'' (where it has over several thousand entries -- including variations such as cartooning and mirroring -- and appeared as a front page Google doodle on December 27, 2013, and even more recently as the cover of Giornale di Fisica, Vol.~62, April 2021, which is addressed to high school physics teachers), is in the possession of the Kremsmünster Benedictine monastery in Austria. In the earliest citation we know \cite{wolf}, the acquisition of the painting is described as a~sale to the abbot of this monastery in 1864.
A~more complete description is provided in 1898 by Günther \cite{gunther}. He states that: 
	``According to the notes I owe to P.~Hugo Schmid, the monastery librarian there, the painting belonged to a~notary Grüner, who passed it on in 1864 to the current abbot of the monastery, Reslhuber.''
The painting was described as a~copy of an original in possession of the descendants of Kepler’s siblings \cite{wolf}. Neither the artist nor the identification of the presumed original were provided in any of the sources. The current painting is an oil on oak panel ($37{\times}50\,$cm) with no signature or attribution. There is only the Latin phrase
	``\textit{Aetatis Suae 39, 1610}''
in the upper right corner (which usually appears cropped in the photographs). The painting was discussed in the 1930 \textit{Kepler Festshrift} by Zinner in a~general summary of Kepler iconography \cite{zinner}. He notes that the painting was sold to the abbey for 200 gulden and that Grüner was from Weil der Stadt (Kepler's birthplace). Zinner also summarizes the previous descriptions, calls the painting an ``alleged portrait'', and notably includes the opinion of experts who examined the painting in the 1920s. Although there is no published physical analysis, Zinner reports that Professor Seraphin Maurer of the Vienna Academy of Fine Arts stated that although this picture gives the the overall impression to be from the 17$^{\rm th}$ century,
	``[h]owever, on closer inspection, the technical treatment shows that the painter had no understanding of portraiture, [\dots]. Furthermore, the colors have not yet faded, which is always the case with pictures from that time. There are no visible signs of aging such as cracks, so I assume that the picture might have appeared around 1800 (a little earlier or later). The oak panel also has a~finish that is not the usual type from around 1600. These are the main features that enable me to state that the picture is a~copy.''
Even before reading Zinner \cite{zinner}, we had long suspected that the painting could not be from earlier than the 19$^{\rm th}$ century for stylistic reasons, and we were quite pleased to find that Zinner and his informants reached the same conclusions.

More to the point, the portrait in Fig.~\ref{fig:kepler}a is likely not even Kepler but we will argue that it is a~19$^{\rm th}$ century forgery that could be based loosely on an official academic portrait from life of Michael Mästlin (see Fig.~\ref{fig:kepler}b), Kepler's teacher and promoter. It is enough to compare the details in the two portraits.

We contend that this alleged painting of Kepler, if it is based on anything other than fantasy, could easily be a~corruption of the Mästlin portrait. Anyone knowing Kepler's dates could compose the inscription. The dress is wrong for the period and does not accord with the two certain historically flanking portraits of Kepler -- the commemorative medallion from his wedding in 1597 and his official portrait from 1620 (see Fig.~\ref{fig:kepler}c). In neither does he wear an academic gown. Although certainly not determinant, it would be somewhat surprising for Kepler at that time in the court to wear the formal collar worn by noblemen and professors of the period, as in the Mästlin portrait. A lace collar (as in Kepler's official portrait) would have been appropriate. An alternate candidate for the presumed original subject could be Wilhelm Schickard who was a~contemporary and a~collaborator of Kepler's but in 1610, he had yet to enter the academic world, so there would be no reason for the portrait.

There is another presumed portrait of Kepler, from around 1610, which since the 1973 has been attributed to Hans von Aachen, one of the favorite painters of Rudolf~II and a~contemporary of Kepler in Prague \cite{kaufmann,iau,jansova}, see Fig.~\ref{fig:kepler}d.  These two portraits cannot be simultaneously the representation of the same person. Although the identification is still disputed, at least in the case of von Aachen, the artist is known and the painting is original.
Finally, another painting identified as Kepler, known as ``the Linz portrait'' \cite{ulm}, is dated to 1620, see Fig.~\ref{fig:kepler}e. Although the artist is unknown, as already noted in the literature, it does resemble the representation of Kepler in the frontispiece of the contemporaneous Rudolphine Tables \cite{albinus}, and Kepler's official portrait from the same year, but here Kepler wears the Spanish collar. This is, however, a miniature \cite{ulm}, \emph{not} an official portrait, and the fashion was much more common in Europe at that time.

So our final question is how did this contagion spread? Before 2000, we cannot find any examples of the portrait being used to represent Kepler, except for our cited references. However, 2001 saw the founding of Wikipedia and the portrait appeared there for the first time in 2005. Thereafter, it becomes ubiquitous. For example, it appears in the ESA press release from 2011 (explicitly citing Wikipedia), ESO attached it to an article from 2016, and NASA used it in their Solar System educational material in 2017.

While this may just seem like a~trivial byway, images fix in the mind. Kepler deserves better.

\bigskip

% Reference
\renewcommand{\refname}{\normalsize References}
{\footnotesize
\bibliographystyle{aip}
\bibliography{kepler-portrait_arXiv_shore-pavlik}}

\subsection*{Note after acceptance}

After placing the preprint on arXiv, we were contacted by Dr.~Alena Šolcová who pointed out several typos in the original arXiv version and suggested that we also include the Linz miniature.

We were also fortunate to have been contacted by Dr.~Anders Nyholm who very kindly provided reproductions of two notices of this painting predating the earliest we knew. One is the catalogue entry for the Vienna World's Fair (Welt-Ausstellung 1873 in Wien, p.~821, group~14 -- Scientific Instruments, no.~3, \url{https://sachsen.digital/werkansicht?tx_dlf[id]=17012\&tx_dlf[highlight_word]=Kepler\&tx_dlf[page]=835}), where the the source is listed as ``\textit{Resslhuber Abt Dr., Kremsmünster. Portr\"at: Johanes Kepler (1610).}'' (spelled with two `s' and one `n') and no further information.
The second citation is an aesthetic appreciation from the Neue Freie Presse (Wien, Donnerstag, October 2, 1873, p.~4, Nature and Ethnology, \url{https://anno.onb.ac.at/cgi-content/anno?aid=nfp\&datum=18731002\&query=\%22Kepler\%22\&ref=anno-search\&seite=20}). It is noted that Reslhuber provided the portrait. It appears that he was convinced that the portrait is real and was proud of the acquisition.

\end{multicols}

\end{document}